\documentclass[doublecol]{epl2}

\usepackage{graphicx}
\usepackage{amsmath,amssymb,amsfonts,color}

\newcommand{\td}{\mathrm{d}}
\newcommand{\ee}[1]{\text{e}^{#1}}

\newcommand{\mc}{\text{ce}}
\newcommand{\ms}{\text{se}}
\renewcommand{\vec}[1]{\boldsymbol{#1}}
\newcommand{\Db}{\bar D}

\begin{document}

\title{Effective Perrin theory for the anisotropic diffusion of a strongly hindered rod}
\date{\today}

\author{T. Munk \and F. H\"of\/ling \and E. Frey \and T. Franosch}
\institute{Arnold Sommerfeld Center for Theoretical Physics (ASC)  and Center for
   NanoScience (CeNS), Fakult{\"a}t f{\"u}r Physik,
   Ludwig-Maximilians-Universit{\"a}t M{\"u}nchen, Theresienstra{\ss}e 37,
   80333 M{\"u}nchen, Germany}

\pacs{05.20.-y}{Classical statistical mechanics}
\pacs{61.20.Lc}{Structure of liquids: Time-dependent properties; relaxation}

\abstract{
  Slender rods in concentrated suspensions constitute strongly interacting
  systems with rich dynamics: transport slows down drastically and the
  anisotropy of the motion becomes arbitrarily large. We develop a mesoscopic
  description of the dynamics down to the length scale of the interparticle
  distance. Our theory is based on the exact solution of the Smoluchowski-Perrin equation;
  it is in quantitative agreement with extensive Brownian dynamics
  simulations in the dense regime. In particular, we show that the tube
  confinement is characterised by a power law decay of the intermediate
  scattering function with exponent 1/2.
}

\maketitle

Brownian motion of highly anisotropic particles is considerably more complex
than the diffusion of spherical objects,
the basic understanding of which is founded on
the seminal works by Einstein and von Smoluchowski.
A shape anisotropy results
in diffusion coefficients that depend on the direction of motion in the body
frame, thus inducing a coupling of translation to the orientation. This
anisotropic dynamics has been investigated in recent experiments measuring
diffusion coefficients of micrometre sized ellipsoids and rods by single
particle tracking \cite{Han:06,Mukhija:07,bhaduri:08}; in particular,
non-Gaussian statistics has been observed~\cite{Han:06}. Likewise in dynamic
light scattering, the rotational and translational diffusion coefficients were
determined simultaneously~\cite{Cush:04}. For these dilute systems, the ratio of
diffusion parallel and perpendicular to the long symmetry axis was limited to
values up to $D_\parallel/D_\perp\approx 4$ in quasi two-dimensional (2D)
confinement.

Considerably higher values of this ratio have been observed in simulations of
semi-dilute suspensions of slender rods, yielding $D_\parallel/D_\perp$ up to
values of 50~\cite{cobb:05,Bitsanis:90}. This increase in anisotropy is caused
by the steric constraints imposed by surrounding rods; thereby the transverse
and rotational motion is suppressed, whereas the longitudinal transport is
barely influenced~\cite{Szamel:93}. An intermediate regime of anisotropic
diffusion has been derived for ballistic needles within kinetic
theory~\cite{otto:06} and was observed in simulations~\cite{hoefling:08c}.

For a finite width $b$, the rods undergo a phase transition to the nematic phase
at densities of the order of $1/b L^2$ as predicted by
Onsager~\cite{onsager:49}. The dynamics in this ordered phase becomes trivially
anisotropic and splits into a fast diffusion along the nematic director axis and
slower diffusion perpendicular to it. Such a pronounced anisotropic diffusion
has been observed in simulations of nematic elongated ellipsoids \cite{Allen:90}
and spherocylinders \cite{Loewen:99,Kirchhoff:96}. Experiments have also clearly
demonstrated orientation-dependent diffusion in colloidal nanorods in the
isotropic and nematic phase~\cite{vanBruggen:98} and in various liquid
crystalline phases of \emph{fd} viruses~\cite{Lettinga:05,Lettinga:07}. The
phenomena connected with the nematic phase transition are essentially 
understood and will not be discussed in this work.

In the isotropic phase, experiments and computer simulations have been
restricted to determine the diffusion coefficients from the mean-square
displacements. The complex interplay of translational and rotational motion as
exemplified for a single free ellipsoid~\cite{Han:06} has not been studied for the
strongly hindered motion in solution yet. In principle, one should
characterise the dynamics in terms of an intermediate scattering function or a van
Hove correlation function, as has been done recently for the smectic
phase~\cite{Lettinga:07}.

A general theory for the anisotropic motion of rods in entangled suspensions is
a long-standing problem, due to the intricacy of the many-body interaction. In
particular, such a theory should include memory effects from the
translation-rotation coupling and explain the emergence of new macroscopic time
and length scales. To account for phenomena which depend on the simultaneous
interaction with many particles, a non-perturbative approach is required.
Substantial progress would be achieved in terms of an effective one-particle
theory that allows for quantitative predictions.

\section{Unhindered motion}

Already at the level of a \emph{single, free} rod the problem is involved: the
dynamics of the probability distribution $\Psi(\vec{R},\vec u, t)$ is governed
by the Smoluchowski-Perrin (SP) equation \cite{Perrin:36,berne:76,doi:86},
\begin{align}
  \label{eq:smol-perr}
  \partial_t \Psi &= -D_\text{rot} \hat L^2 \Psi +
  \partial_{\vec{R}} \cdot \left[ (D_\parallel-D_\perp) \vec{u} \,\vec{u}
    + D_\perp \mathbb{I} \right] \cdot
  \partial_{\vec{R}} \Psi,
\end{align}
with the centre of mass position of the rod $\vec R$, its orientational unit
vector $\vec u$, the rotational diffusion coefficient $D_\text{rot}$, and the
angular part of the Laplacian, $-\hat L^2$. The full formal solution of this
equation in 3D was given in ref.~\cite{aragon:85}; its quantitative evaluation
is still missing, although certain aspects are well
understood~\cite{berne:76,doi:86}. To the best of our knowledge, the solution
$\Psi(\vec R,\vec u,t)$ of the Smoluchowski-Perrin equation has not been
discussed for 2D systems. Equation~\eqref{eq:smol-perr} is trivially solved for
isotropic diffusion, $D_\perp=D_\parallel$, and at macroscopic time scales,
$t\gg 1/D_\text{rot}$, where the translation-rotation coupling is relaxed. Then,
the second term may be averaged over $\vec u$, yielding the average diffusion
coefficient.

\section{Model for interacting rods}

We study a model for the sterically constrained dynamics of rods
that naturally induces anisotropic diffusion of arbitrarily large ratios
$D_\parallel/D_\perp$. The model is set up in a 2D environment; the dynamic
  properties we focus on are however equally present in a 3D embedding space.
We demonstrate that the Smo\-lu\-chow\-ski-Perrin equation provides an excellent
effective theory for the dynamics, when the measured diffusion coefficients
serve as input parameters. In particular, we compare mean-square displacements
(MSDs) and the intermediate scattering function from computer simulations with
their exact results from eq.~\eqref{eq:smol-perr}. Furthermore, we have
discovered an intermediate algebraic decay in the intermediate scattering
function, characteristic for the anisotropic sliding motion.

The model considers the overdamped motion of a single rod with zero width
exploring a plane with randomly distributed, hard point obstacles \cite{hoefling:08}. Then the
orientational unit vector of the rod is parametrised by one angle, $\vec u(t)
=\boldsymbol(\!\cos\vartheta(t), \sin\vartheta(t)\boldsymbol)$, and $\hat L^{2}$ reduces to
$-\partial_{\vartheta}^2$. The model ignores any excluded volume to exclusively
concentrate on entanglement effects. The physical properties of this system are
thus controlled by a single parameter, the reduced density $n^{*}:=nL^{2}$,
where $L$ denotes the length of the rod, and $n$ the number density of the
obstacles. The model is closely related to 3D suspensions of rods when
considering a planar section. A tagged rod found initially in this plane is
approximately confined to it by the neighbouring rods for times shorter than the
orientational relaxation time, $t\ll\tau_\text{rot}:=D_\text{rot}^{-1}$, the
largest time scale present in the system \cite{hoefling:08}. The constrained
motion of the rod then corresponds to diffusion in a planar course of localised
intersection points.

\begin{figure}[bt]
  \includegraphics[width=\columnwidth]{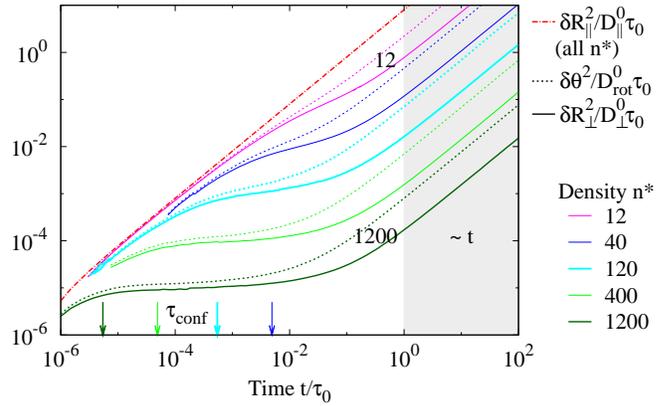}
  \caption{(colour online) Simulated body-frame MSD in the entangled regime. For
    each density, three observables are shown: The MSD parallel to the rod's
    axis (topmost, dash-dotted), perpendicular (solid), and the MSAD (broken).
    Arrows at the bottom indicate $\tau_\text{conf}$ for the corresponding
    density; the diffusive regime is shaded.}
  \label{fig:msd}
\end{figure}

\begin{figure}[bt]
  \begin{center}
    \includegraphics[width=\columnwidth]{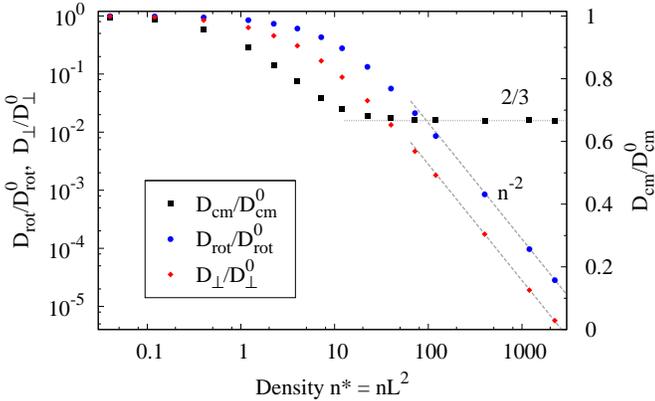}
  \end{center}
  \caption{(colour online) Density dependence of the diffusion coefficients. Left
    axis: rotational diffusion and transverse diffusion; right axis: centre of
    mass diffusion.}
  \label{fig:diffconst}
\end{figure}

\section{Anisotropic diffusion}

The microscopic motion of the rod is diffusive, with diffusion coefficients
chosen according to first order hydrodynamics of a slender rod,
$D^{0}_\perp=D_{\parallel}^{0}/2$, and
$D^{0}_\text{rot}=6D_{\parallel}^{0}/L^2$. For the computer simulations we have
combined the Langevin equations corresponding to eq.~(\ref{eq:smol-perr}) with
an event-driven algorithm to detect the collisions between rod and obstacles
\cite{Tao:06,hoefling:08,hoefling:08b}. The MSDs in fig.~\ref{fig:msd} visualise
essential properties of the model in the semi-dilute regime, $n^{*} \gg 1$. We
define the displacement in the \emph{body-fixed} frame along the axis as $\Delta
R_\parallel(t) := \int_0^t \dot{\vec{R}}(t') \cdot \vec{u}(t') \td t'$, and
similarly the transverse part $\Delta R_\perp(t)$. The parallel MSD, $\delta
r_{\parallel}^{2}(t) :=\langle \Delta R_{\parallel}(t)^{2}\rangle$, is not
affected by the obstacles at all, due to zero excluded volume. Consequently, the
parallel diffusion coefficient is independent of the density,
$D_{\parallel}\equiv D_{\parallel}^0$. In contrast, the perpendicular MSD
$\delta r_\perp^2(t)$ and the mean-square angular displacement (MSAD) $\delta
\vartheta^2(t)$ enter a plateau beyond a density-dependent time scale
$\tau_\text{conf}$. The plateau reflects the local confinement to an effective
cage built up by the surrounding obstacles, referred to as
``tube''~\cite{doi:86}. Its diameter $d$ is determined by $nLd\approx 1$,
leading to a relation for the time when the confinement becomes effective,
$\tau_\text{conf}:=d^{2}/D_{\perp}^{0}\approx 1/ n^2 L^{2} D_{\perp}^{0}$. At
the time scale $\tau_{0}=L^{2}/D_{\parallel}$, the rod moves a distance
comparable to its length $L$, hence it leaves the tube and the MSDs become
diffusive again. From the long-time asymptotes, the diffusion coefficients are
read off, e.g., $D_{\perp}(n^{*})= \lim_{t\to\infty} \partial_t \delta
r_{\perp}^{2}(t)/2 $. Fig.~\ref{fig:diffconst} demonstrates the huge suppression
of perpendicular and rotational diffusion coefficients: both scale with obstacle
density as ${n}^{-2}$, as has been argued by Szamel~\cite{Szamel:93}. A
  consequence of the vanishing perpendicular component is the saturation of the
  centre of mass diffusion,
  $D_\text{cm}(n^{*}\gg 1)
  =2D_\text{cm}^{0}/3$. For the largest simulated density, $n^*=2240$, the
anisotropy ratio approaches a value of $D_\parallel/D_\perp\approx
10^{5}$. Note
that these large ratios in dense systems do not require an anisotropy on the
microscale; rather the anisotropic motion is generated dynamically from the
strong interaction with many obstacles.

\begin{figure}[bt]
  \includegraphics[width=\columnwidth]{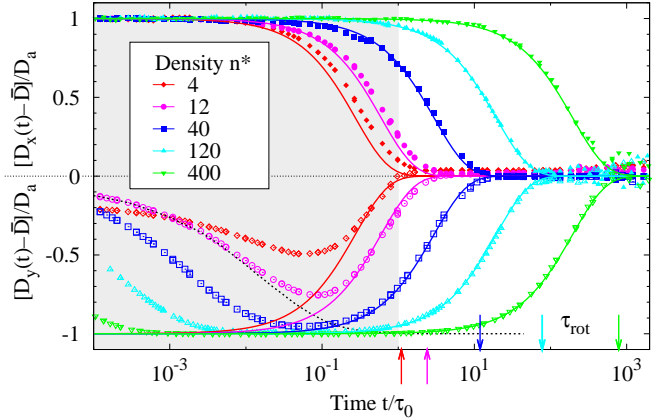}
  \caption{(colour online) Deviations of the diffusion coefficients $D_x(t)$ and
    $D_y(t)$ in the space-fixed frame from the isotropic value $\Db$. The
    initial orientation is fixed to the $x$-axis; normalisation is chosen such
    that values of $\pm 1$ indicate diffusion with $D_{\parallel}$ and
    $D_{\perp}$, respectively. Symbols show results of simulations, solid lines
    the effective Perrin theory; arrows indicate $\tau_\text{rot}$ for the
    different densities. The broken line displays the measured $[ D_\perp(t)
    -\Db]/D_a$ at $n^* =12$ for comparison. For $t\to 0$, the
      perpendicular diffusion in the simulations asymptotically approaches 
      $[D_{y}(t\to 0)-\Db]/D_{a}= [1-D_\parallel/D_{\perp}(n^*)]^{-1}$.}
  \label{fig:msdfxd}
\end{figure}

A characteristic crossover between aniso\-tropic and isotropic dynamics is seen
clearly when plotting the time-dependent diffusion coefficients in a
\emph{space-fixed} frame, $D_x(t) := \partial_t \langle\Delta
x(t)^2\rangle_{\text{f}}/2$ and $D_y(t) := \partial_t \langle\Delta
y(t)^2\rangle_{\text{f}}/2$, where the subscript `$\text{f}$' indicates that the
initial orientation is fixed to the $x$-axis; seen fig.~\ref{fig:msdfxd}. The
memory of the orientation is lost only at times larger than
$\tau_{\text{rot}}\sim {n}^{2}$, resulting in a time window of anisotropic
diffusion that is significantly prolonged with increasing density.
Note that fig.~\ref{fig:diffconst} directly visualises the density
  dependence of the timescale indicating the crossover to isotropic dynamics,
  $\tau_\text{rot}\equiv D_\text{rot}^{-1}$.

For comparison with the unhindered anisotropic motion, eq.~(\ref{eq:smol-perr})
is solved for the conditional probability distribution $\Psi(\vec{R},\vartheta,
t|\vartheta_0)$, with the initial condition
$\Psi(\vec{R},\vartheta,t=0|\vartheta_0) = \delta(\vec{R})
\delta(\vartheta-\vartheta_0)$. A Fourier transform defines the characteristic
function $ G_{\vec{k}}(\vartheta, t|\vartheta_0) = \int\! \text{e}^{-\text{i}
  \vec{k} \cdot \vec{R}}\, \Psi(\vec{R}, \vartheta, t|\vartheta_0) \td\vec{R}$.
Its equation of motion attains the form of a Schr{\"o}dinger equation,
\begin{equation}\label{eq:schr}
 \partial_t G_{\vec{k}} = - \hat{\cal H}_0 G_{\vec{k}} - \hat{V} G_{\vec{k}},
\end{equation}
with the operators $ \hat{\cal H}_0 = -D_\text{rot} \partial_\vartheta^2 $ and
$\hat{V} = (D_\parallel-D_\perp) (\vec{k} \cdot \vec{u} )^2 + D_\perp k^2$.
Perturbation theory in $kL\ll 1$ now solves iteratively for $G_{\vec k}$. The
central quantity of interest is the intermediate scattering function, 
$F(\vec{k},t |\vartheta_0) := \int G_{\vec{k}}(\vartheta, t |
\vartheta_0)\td\vartheta$. Up to fourth order
in $k$,
\begin{align}\allowdisplaybreaks
    \lefteqn{F(\vec{k}, t| \vartheta_0) = 1 - \Db k^2 t-\frac{D_a}{2} \tau_4(t)
    \left(k_{+}^{2} 
         +k_{-}^{2} \right)} \notag\\
    &\quad +\frac{\Db^2}{2}k^4t^2+\frac{\Db D_a}{2}k^2t\tau_4(t) \left(
      k_+^2+k_-^2
    \right)  \label{eq:86}\\\notag
    &\quad +\frac{D_a^2}{8D_\text{rot}} \left\{ \frac{\tau_4(t)-\tau_{16}(t)}{6}  \left(
        k_+^4+k_-^4\right)
     +k^4 [t-\tau_4(t)] \right\}.
\end{align}
The notation is abbreviated with the isotropic and anisotropic diffusion
coefficients, $ \Db = (D_\parallel+ D_\perp)/{2}$, $ D_a =
(D_\parallel-D_\perp)/{2}$, the wavevector components $k_{\pm}=(k_{x}\pm
ik_{y}) \ee{\mp i \vartheta_0}$, and $\tau_j(t) := \int_0^t \text{e}^{- j
  D_\text{rot} s} \td s$. Note that the dependence of eq.~\eqref{eq:86}
on $\vartheta_0$ is hidden in the definition of $k_\pm$---this property originates
from rotational symmetry and holds to all orders in $k$. From $F(\vec{k}, t|
\vartheta_0)$ all moments are obtained by derivatives, e.\,g., $\langle \Delta
x(t)^2\rangle_{\text{f}}=-\partial_{k_x}^2F(\vec{k}, t| \vartheta_0)|_{\vec
  k=0}$. Some of these moments have been calculated by Han et al.\ \cite{Han:06}
from the Langevin equations equivalent to eq.~(\ref{eq:smol-perr}).

\section{Effective Perrin theory}

The idea is to use the solutions of the Smoluchowski-Perrin equation
with the diffusion coefficients $D_\parallel, D_\perp(n^*),
  D_{\text{rot}}(n^*)$---measured in the simulations---to obtain a prediction for
  time-dependent MSDs and the intermediate scattering function in the presence of
  obstacles. In other words, the results from the simulations for two transport
  coefficients in semi-dilute systems are employed in the theory derived for
  \emph{free} diffusion to describe the full time-dependence of a
  \emph{semi-dilute} suspension. Since the orientation changes only
  gradually, the long-time rotational motion is described by
  diffusion~\cite{hoefling:08}. Thus the description in terms of an effective
  Perrin theory has to be trivially correct in the relaxed regime where
  translational diffusion is isotropic, i.e., on macroscopic time and length
  scales, $t\gtrsim \tau_\text{rot}$ and $k^{-1} \gtrsim L_\text{rot} :=
  \sqrt{D_a \tau_\text{rot}}$. 

We will show that in fact the effective Perrin theory constitutes a
  quantitative \emph{mesoscopic} theory in the dense regime, $n^* \gg 1$, and
  successfully describes also the translation-rotation coupling induced by the
  interaction with the many obstacles. That such an approach should work in
principle has been anticipated earlier for dense needle liquids~\cite{otto:06}.
A comparison with simulated MSDs, fig.~\ref{fig:msdfxd}, reveals excellent
agreement down to the time scale $\tau_0$. For short times, $t\ll \tau_0$, the
space- and body-fixed frames coincide, implying $D_x(t) \simeq D_\parallel(t)$
and $D_y(t) \simeq D_\perp(t)$.

To access the full range of wavenumbers, we construct the exact solution of
$G_{\vec k}(\vartheta,t|\vartheta_0)$ in terms of Mathieu functions. The
equation of motion, eq.~\eqref{eq:schr}, is rewritten as
\begin{equation}
  \partial_t G_{\vec{k}} = D_\text{rot} \partial^2_{\vartheta} G_{\vec{k}}-k^2
  \left(
    \Db+D_a\cos 2\vartheta
  \right)G_{\vec{k}},\label{eq:23}
\end{equation}
in a coordinate frame with $\vec k=k\hat e_{x}$. A separation ansatz, $
G_{\vec{k}}(\vartheta, t) = g_{\vec{k}}(\vartheta)\ee{-\lambda t}$, yields the
Mathieu equation, 
$ 0 = \partial^2_\vartheta g_{\vec{k}}+
\left(
  a-2q\cos 2\vartheta
\right)g_{\vec{k}}, $ 
with the parameter $q=D_ak^2/2D_\text{rot}$ and the eigenvalue $a=(\lambda
-\Db k^2)/{D_\text{rot}}$. 
The general solution is thus a linear combination of even and odd
eigenfunctions, $\mc_{j}(\vartheta,q)$ and
$\ms_{j}(\vartheta,q)$~\cite{Abramowitz:84};\footnote{We use the normalisation 
  $\int_0^{2\pi}\mc_{j}^{2}(\vartheta,q)\td\vartheta=1$ and
  $\int_0^{2\pi}\ms_{j}^{2}(\vartheta,q)\td\vartheta=1$.} the decay rate
$\lambda=\lambda(a,k)=aD_\text{rot}+\Db k^2$ depends on the corresponding even
and odd eigenvalues, $a\to a_{j}(q)$ and $a\to b_{j}(q)$, respectively:
\begin{multline}
  \label{eq:69}
    G_{\vec k}(\vartheta, t|\vartheta_0)=\sum_{j=0}^\infty
  \left[
    \ee{-\lambda(a_j,k)t}\mc_j(\vartheta_0,q)\mc_j(\vartheta,q)\right.\\
    +\left.\ee{-\lambda(b_j,k)t}\ms_j(\vartheta_0,q)\ms_j(\vartheta,q)
  \right].
\end{multline}
The intermediate scattering function with unconstrained initial orientation is
obtained by integrating over $\vartheta$ and $\vartheta_0$,
\begin{equation}
  \label{eq:scat-fct}
    F(\vec k,t) =
    \ee{-k^2 \Db t}\sum_{j=0}^\infty \ee{-a_{2j}(q)D_\text{rot}t}
    \left[
      A_{0}^{(2j)}(q)
    \right]^{2},
\end{equation}
with  coefficients $A_{0}^{(2j)}(q):= 
\int_0^{2\pi}\!\mc_{2j}(\vartheta,q)\td\vartheta/\sqrt{2\pi}$.  

The convergence of the sum is determined by the magnitude of $q$. The
eigenvalues $a_j(q)$ are ordered ascendingly in $j$; furthermore, the
coefficients fulfil $A_{0}^{(2j)}(q) = {\cal O}(q^j)$. Hence for $q\ll 1$, the
low-$j$ terms yield the major contributions to the sum, $F(\vec k,t) = \ee{-k^2
  \Db t} \left[\ee{q^2 D_\text{rot} t/2 }(1-q^2/8) + \ee{-4 D_\text{rot} t}
  q^2/8 + {\cal O}(q^4)\right]$, representing the first correction to isotropic
diffusion. Thus $q^2 \lesssim 1$ defines the macroscopic regime, corresponding
to $ (k L_\text{rot})^4 \lesssim 4$, and $L_\text{rot}$ is derived as the
relevant macroscopic scale.

The opposite limit, $q\gg 1$, is relevant for large densities due to the
suppression of $D_\text{rot}$. The asymptotic expansion 
$a_j(q) \simeq -2 q + (4j+2) \sqrt{q} + {\cal O}(1)$ reveals a reduction of the
exponential prefactor in eq.~(\ref{eq:scat-fct}) to $\ee{- k^2 D_\perp t}$, and
a large number of terms contribute. Then, the terminal relaxation is ruled by an
exponential with decay rate $\tau_\text{term}^{-1}:=k\sqrt{2 D_{a}D_\text{rot}}+
k^2 D_\perp $.

\begin{figure}[!b]
  \includegraphics[width=\columnwidth]{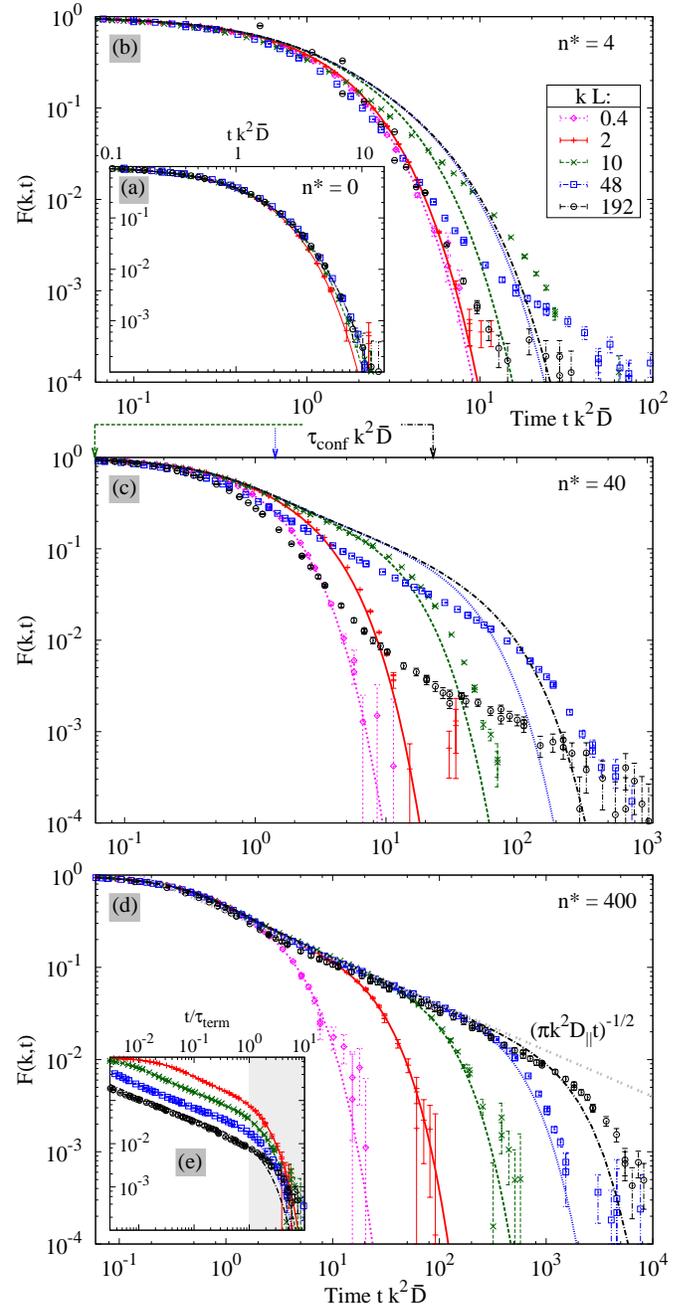}
  \caption{(colour online) Time and wavenumber dependence of the intermediate scattering function for
    densities (a) $n^{*}=0$, (b) $n^{*}=4$, (c) $n^{*}=40$, and (d), (e) $n^{*}=400$.
    Symbols represent simulation results, lines the effective Perrin theory,
    eq.~(\ref{eq:scat-fct}), and time is in units of $k^2 \Db$.
    By rescaling time with $\tau_\text{term}$, inset (e) visualises the 
    terminal relaxation.}
  \label{fig:scat}
\end{figure}

\begin{figure}
  \includegraphics[width=\columnwidth]{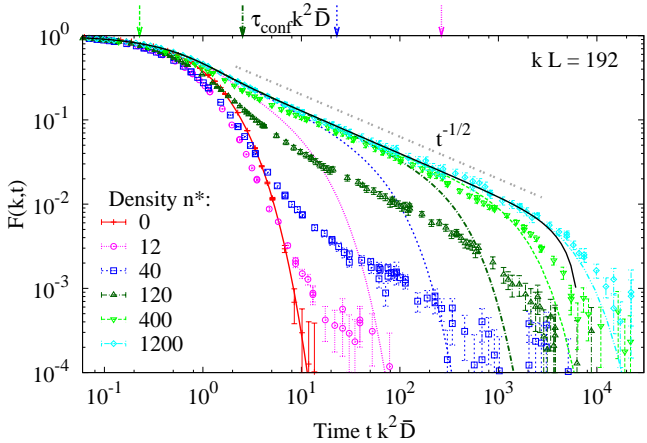}
  \caption{(colour online) intermediate scattering function for fixed wavenumber
    and varying density. Increasing confinement is manifested in the simulations
    (symbols) by the development of a power law decay. Lines show the effective
    Perrin theory; the perturbative correction to eq.~\eqref{eq:order0} is
    indicated by the thick black line for $n^*=1200$.}
  \label{fig:scatk}
\end{figure}

Based on the simulation results for the intermediate scattering function, we
test the range of validity of the effective Perrin theory in
fig.~\ref{fig:scat}. The quality of our data, approaching a signal-to-noise
ratio of $10^{-4}$ in the scattering function, allows for a clear distinction of
features on a large range of time and length scales.\footnote{To achieve
  sufficient statistics, we collected at least 350 trajectories for each
  density. For the largest densities, the simulation of a single trajectory took
  about 13 days of CPU time on a AMD Opteron\textsuperscript\textregistered\ 2.6
  GHz core.} By construction, the theory describes the data at zero density
[panel~(a)].
For intermediate densities, $n^{*} \simeq 1$, the Perrin prediction
  strongly deviates from the simulation. The two lowest wave numbers are in the
  macroscopic regime; correspondingly the curves coincide after proper
  rescaling [panel (b)], and the Perrin prediction applies  in the trivial sense.
 At larger densities, the agreement becomes increasingly accurate as
the wave number decreases [panels~(c), and (d)] also for wavenumbers beyond the
macroscopic regime. 
 Second, deviations are shifted
 to larger wavenumbers for higher densities. 
We conclude that there is a
density-dependent length scale $\xi$ determining the validity of our
coarse-grained approach. From all simulated densities, we have identified $\xi$
as the typical distance between obstacles, $\xi := n^{-1/2}$: for the shown
densities, $n^*=\{4,40,400\}$, the corresponding wavenumbers $2\pi/\xi$ are
$13 L^{-1}$, $40 L^{-1}$ and $126 L^{-1}$, respectively. The slowing down of the dynamics
with increasing confinement is displayed in fig.~\ref{fig:scatk}. For high
wavenumbers $k\sim 2\pi/d$ the intermediate scattering function probes the
formation of the tube at time scales $\tau_{\text{conf}}$, see
figs.~\ref{fig:scat}c and \ref{fig:scatk}. Once the tube confinement becomes
effective, an intermediate algebraic decay emerges, $F(\vec k,t)\sim t^{-1/2}$,
which we attribute to the sliding motion inside the tube. This power law is cut
off by an exponential relaxation at $\tau_\text{term}$, see
fig.~\ref{fig:scat}e.

\section{Origin of the power law}

The power law is hidden in eq.~\eqref{eq:scat-fct} in the sum of many
exponentials for $q\gg 1$. For strongly suppressed perpendicular and rotational
motion, eq.~(\ref{eq:23}) is approximated by $\partial_t G_{\vec{k}} = -k^2
D_\parallel\cos^2(\vartheta)\, G_{\vec{k}}$, which yields
\begin{equation}
  \label{eq:order0}
  F(\vec k,t) = \ee{-k^{2}D_{\parallel}t/2}I_{0}\big(k^{2}D_{\parallel}t/2\big) 
  \simeq \left(\pi k^{2}D_{\parallel}t\right)^{-1/2}.
\end{equation}
The second relation results from an expansion of the modified Bessel function of
the first kind $I_{0}(z)$ for large argument; it quantitatively reproduces the
scattering function in the power law regime, shown in fig.~\ref{fig:scat}d.
Perturbation theory in $D_\text{rot}$ also captures the terminal relaxation with
the previously calculated $\tau_\text{term}$, see fig.~\ref{fig:scatk}.

\section{Conclusions}

From our analysis we conclude that the motion of thin rods in concentrated
suspensions exhibits a rich interplay of time and length scales, as exemplified
in the MSDs and intermediate scattering function. The hindered motion leads to a
strong anisotropic dynamics manifested in a significant translation-rotation
coupling which persists up to macroscopic length and time scales, $L_\text{rot}$
and $\tau_{\text{rot}}$. This dynamically induced coupling is an emergent
phenomenon with long memory effects. The confinement in the tube gives rise to a
well separated spectrum of length scales $d \ll \xi \ll L \ll L_\text{rot}$. The
tube becomes effective once the rod encounters new steric constraints---thus the
interparticle distance $\xi$ constitutes the lower length scale of a mesoscopic
window, where an effective theory with renormalized parameters becomes valid;
this regime extends up to $L_\text{rot}$. Note that such a window opens only in
the strongly anisotropic regime and is absent for, e.g., ellipsoids with moderate
aspect ratio. There the orientational motion is dominated by excluded volume
effects, and becomes neither diffusive nor
exponential~\cite{kaemmerer:97,Pfleiderer:08}. For strongly entangled
suspensions, the orientation changes only gradually by tube renewals, and it
has to be included in the set of slow degrees of freedom in addition to the
translation. The effective Perrin theory constitutes a Markov process in these
variables, and the long memory observed in the intermediate scattering function
is generated by integrating out the slowly varying orientation. This mesoscopic
description is independent of the details of the tube generation; in
particular, the obstacles may also fluctuate in time and space, or even disappear.

The Perrin approach fails to capture the fact that strongly confined needles
have to diffuse along their axis a distance $L$ to relax the tube constraint.
However, since the rotational diffusion $D_\text{rot}$ is much slower than
$D_\parallel /L^2$, this appears to be negligible, at least on the scales
investigated here.

We expect that our findings are also relevant for slender rods of
  finite width $w$. Then the motion of a single rod is again strongly confined
  by a tube comprised of the surrounding particles. For large aspect ratios
  $L/w$ the isotropic-nematic spinodal occurs at 3D concentrations $c\approx
  4/wL^{2}$~\cite{Tao:06}, e.g., at $c=200/L^{3}$ for $L/w=50$. For densities
  $100\leq cL^{3}\leq 150$, 3D simulations of suspensions of rods with this
  aspect ratio exhibit significantly anisotropic diffusion of about $25\leq
  D_\parallel/D_{\perp}\leq 50$ \cite{Bitsanis:90,cobb:05}. In our 2D model such
  ratios are attained for $n^{*}\approx 20$, already at the onset of the regime
  of entangled diffusion. Experimental realisations of highly entangled rod
  suspensions may be achieved, e.g., for the tobacco mosaic virus. Mutants of
  this rod-shaped virus have been observed with an aspect ratio of 50 and
  higher~\cite{Miller:07}, thus entangled suspensions below the nematic phase
  transition should be feasible. Another promising model system are solutions of
  microtubules; their aspect ratio can even be considerably
  larger~\cite{Gittes:93,pampaloni:06}, and with a ratio of 100 they can still
  be considered as approximately stiff rods. A third example are solutions of
  \emph{fd} viruses, exhibiting a high degree of monodispersity and a comparable
  aspect ratio of about 130~\cite{Lettinga:07}.

In the preceding discussion we have neglected hydrodynamic interactions
  mediated by the solvent. Inclusion of these effects present a considerable
  challenge both to the theory and the simulation techniques, and is beyond the
  scope of our work. The possible influences of hydrodynamic forces are twofold:
  First, a logarithmic length dependence, $\ln (L/w)$, decorates the microscopic
  expression for the diffusion coefficients. Since the ratios of $D_{\perp}$,
  $D_{\parallel}$, and $D_\text{rot}$, are independent of this correction, this
  effect is already accounted for by choosing $\tau_{0}=L^{2}/D_{\parallel}^{0}$
  as the basic unit of time, as we have done here. Second, nonadditive
  hydrodynamic forces between the rods might change the observed dynamics, in
  particular at short times~\cite{guzowski:08}. Recently, Pryamitsyn and
  Ganesan~\cite{pryamitsyn:08} have argued on the basis of computer simulations
  that effects of hydrodynamic interactions on the translational and rotational
  diffusivity are secondary relative to the steric interactions.
  Assuming that hydrodynamic interactions merely renormalise the macroscopic
  transport coefficients, the mechanism of translation-rotation coupling remains
  unaffected, and our theory should still be applicable.

It is straightforward to extend the concept of an effective Perrin theory to 3D
suspensions of long, thin rods since the motion in the tube is essentially
one-dimensional. The formal solution is then provided in terms of spheroidal
wave functions~\cite{otto:06,aragon:85}, which serves as a starting point to
calculate the intermediate scattering function. In particular, the confined
motion of rods is characterised again by a power law decay of the intermediate
scattering function, which we predict to $F(\vec k,t)\simeq (4 k^2D_\parallel
t/\pi)^{-1/2}$. This algebraic decay constitutes a generic feature of the
sliding motion and should be observable directly in scattering experiments.

\acknowledgments

We thank Matthias Fuchs for stimulating discussions and Annette\ Zippelius for
drawing our attention to ref.~\cite{aragon:85}. Financial support has been
granted by the Nanosystems Initiative Munich (NIM) and by the Deutsche
Forschungsgemeinschaft (DFG) contract number FR\ 850/6-1.


\end{document}